\documentclass{article}
\usepackage{ijcai26}

\usepackage{times}
\usepackage{soul}
\usepackage{url}
\usepackage[hidelinks]{hyperref}
\usepackage[utf8]{inputenc}
\usepackage[small]{caption}
\usepackage{graphicx}
\usepackage{amsmath}
\usepackage{amsthm}
\usepackage{booktabs}
\usepackage{algorithm}
\usepackage{algorithmic}
\usepackage[switch]{lineno}

\title{[COMP25] The Automated Negotiating Agents Competition (ANAC) 2025 Challenges and Results}

\author{
Reyhan Aydo\u{g}an$^{1,2}$
\and
Tim Baarslag$^{3,4}$\and
Tamara C.P. Florijn$^{3,5}$\and
Katsuhide Fujita$^6$\and
Catholijn M. Jonker$^{2,7}$ \and
Yasser Mohammad$^8$ \\
\affiliations
$^1$\"{O}zye\u{g}in \"{U}niversity,
$^2$Delft University of Technology,
$^3$Centrum Wiskunde \& Informatica (CWI),
$^4$Eindhoven University of Technology,
$^5$Utrecht University, 
$^6$Tokyo University of Agriculture and Technology \& National Institute for Advanced Industrial Science and Technology,
$^7$ Leiden University,
$^8$NEC Corporation  \& National Institute for Advanced Industrial Science and Technology\\
\emails
reyhan.aydogan@ozyegin.edu.tr,
T.Baarslag@cwi.nl,
tamara.florijn@cwi.nl,
katfuji@cc.tuat.ac.jp,
C.M.Jonker@tudelft.nl,
y.mohammad@nec.com
}

\begin{document}

\maketitle

\begin{abstract}
   This paper presents the primary research challenges and key findings from the 15th International Automated Negotiating Agents Competition (ANAC 2025), one of the official competitions of IJCAI 2025. We focus on two critical domains: multi-deal negotiations and the development of agents capable of concurrent negotiation within complex supply chain management environments. Furthermore, this work analyzes the results of the competition and outlines strategic directions for future iterations.
\end{abstract}

\section{Introduction}

Since $2010$, the International Automated Negotiating Agents Competition (ANAC) has spearheaded innovation in autonomous agent research, introducing foundational challenges to the multi-agent systems community \cite{jonker2017automated}. Since its inception with 7 participants \cite{ANAC2010}, the competition has expanded significantly, surpassing 850 total participants across 16 editions \cite{ANAC2010-2015}. The 2025 competition featured two distinct leagues. The Automated Negotiation League (ANL) required participants to develop bilateral agents that can engage in sequential multi-deal negotiations with multiple opponents and inter-dependent utility functions. In the Supply Chain Management League (SCML) \cite{yas19scml}, the objective is to design agents that maximize profit within a competitive market. These agents must strategically navigate concurrent negotiations to secure raw materials and sell manufactured goods.

The 2025 edition attracted 142 international researchers forming 42 teams, competing for a 2,500 EUR prize pool. Both leagues utilized the NegMAS \cite{mohammad2021negmas} framework for agent development and tournament execution. Agents competed in a round-robin format across randomized scenarios. ANL performance was evaluated via individual utility and Nash distance, whereas SCML rankings were determined by cumulative profit within the market simulation. The subsequent sections detail the specific league configurations and analyze the final results.

Moreover, ANAC 2025 introduced a pilot competition between IJCAI 2025 participants in which these participants negotiated against autonomous agents in a variety of scenarios for a 1,000 EUR prize pool. The scenarios included a trade negotiation inspired by the Supply Chain Management League, an Island survival tool distribution negotiation and a Grocery items distribution scenario. The participants conducted $1,456$ negotiations and two winners were announced at the closing session of IJCAI 2025. 

\section{Main League Challenges \& Results}

In the main league of 2025, each participating agent encounters multiple opponents in sequence and is rewarded for the specific combination of the deals made in each negotiation. The challenge of this league is \emph{sequential multi-deal} negotiation, where the participants need to coordinate an enormous number of options available and design an agent that outperforms other contestants by conceding efficiently and obtaining the best deals.

At the start of each negotiation, one agent will receive the role \emph{center agent}, while the others receive the role \emph{edge agent}. The center agent negotiates against multiple opponents (the edge agents) in a sequential manner, in what we call \emph{subnegotiations}. Each subnegotiation will follow the Alternating Offers Protocol \cite{AlternatingProtocol} where the starting agent makes an opening offer, which is followed by acceptance, a counteroffer, or a walk-away, repeated in a turn-taking fashion. If the negotiation ends, the center agent continues to the subsequent subnegotiation. 

The agents negotiate over special types of domains that are characterized by the utility function of the center agent, two of which we discuss here. In the \emph{Job Hunt scenario}, a job hunter (center agent) negotiates about two issues: the number of office days and the salary. In the end, the \emph{center utility function} is defined in such a way that he receives the maximum of all the deals that he gathered. His preferences within a subnegotiation (called \emph{side utility function}) are modeled using a linear additive function. Secondly, the \emph{Target Quantity scenario} models a buyer (center agent) that negotiates with multiple sellers over the purchase quantity of a single product. The buyer aims for a specific quantity, for example to buy 10 products i.e.\ a peak preference at 10 products, while buying more or less gradually decreases the utility to 0. 

In total, 17 teams submitted to ANL 2025, of which 12 were selected as finalists. The results of this year are shown in Table~\ref{tbl:anl}. As the negotiation protocol is sequential, the agents need to take into account what future actions will influence their current best decision. When the number of bids and thus the number of possible future actions grows, the number of combined outcomes to consider grows exponentially.

One of the winner agents, RUFL, tackled the computational challenge by constructing a tree to estimate the expected utility. The agent maintains a probability distribution over all the expected values of the child outcomes where their probabilities are generated via a softmax over their expected values with some temperature. By limiting the tree search at a certain level, they keep their computational time within limits. The second winner agent, SAC Agent, is built using the Soft Actor-Critic (SAC) reinforcement learning framework. The SAC bidding policy is a time-dependent concession strategy, where the concession rate is guided by the SAC model.

\begin{table}[ht!]
	\centering
	\caption{ANL 2025 Results}
	\label{tbl:anl}
	\setlength{\tabcolsep}{2pt}
	\begin{tabular}{cllll}
		\toprule
        Rank & Agent & As Center & As Edge & Final Score \\
        \midrule
		
		  1  & RUFL                          &0.714 & 0.084& 0.399\\
		   1  & SAC Agent                          &0.733 & 0.064 & 0.399\\
		   3  & UFunAtAgent                       &0.686 & 0.078& 0.382\\
		% \midrule
		% \multicolumn{6}{l}{TUAT: Tokyo University of Agriculture and Technology}\\
		\bottomrule
	\end{tabular}
\end{table}

Other participants approach the uncertain future deals and their influence on the search space in distinct ways:
\begin{itemize}
    \item Pessimistic. These agents pretend that no further deals will be made (e.g., UfunATagent, SAC, default ANL 2025 strategies).
    \item Contingent. These agents use a probability distribution to assess the likelihood of further deals in the future (e.g., ProbaBot, RUFL, RivAgent, WAgent).%Probabot used uniform distribution, RUFL used a softmax function, and approximated layers that were too deep. WAgent in some cases.
    \item Optimistic. These agents assume that the future will unfold exactly as planned (e.g., WAgent under conditionals).
\end{itemize}
Agents using the pessimistic methods circumvent the computational search challenge by ignoring future deals, disregarding any influence of future deals, limiting their performance. However, using contingent or optimistic approaches takes more computational time. 
%To implement this, some agents used a dynamic target in their strategy (e.g., EOHAgent, CARCAgent2025) or used reinforcement learning techniques (e.g., SacAgent). 
All in all, the biggest challenge turned out to be the memory explosion caused by the big combined outcome spaces (e.g., CARC2025, OzUAgent, KDY, SmartNegotiator, ProbaBot), tackled with different techniques such as a dynamic target (e.g., EOHAgent, CARCAgent2025), sampling methods (e.g., the Memorizer, kAgent), dynamic programming (Astrat3m) or reinforment learning techniques (e.g., SacAgent).

\section{Supply-Chain Management League}
    
The 2025 Supply Chain Management League (SCML) comprised two tracks: OneShot and Standard. A total of $20$ qualified teams participated, with $11$ competing in the OneShot track and $9$ in the Standard track. Because the game mechanics remained identical to SCML 2024, participants were able to leverage the open-source code and technical reports of previous agents. 

To further support the research community, the 2025 edition simplified reinforcement learning and MARL integration by providing a standard Gymnasium \cite{towers2024gymnasium} and a petting-zoo~\cite{pettingzoo} environments that encapsulate the SCML simulation \cite{yasser2024scmlrl}.

The primary objective of SCML is to align automated negotiation research with the practical challenges of industrial applications. The simulation serves as an abstraction of a fundamental procurement dilemma: \textit{How can production needs be met while minimizing the combined costs of procurement and inventory?}
In the SCML OneShot world, multiple autonomous agents manage factories within a supply chain, buying raw materials and selling final products. The market is driven by a set of exogenous un-negotiable contracts for the raw materials and final products. Every simulated days, agents must reach agreements with their suppliers/consumers for buying their production needs and selling their produced items. Contracts specify price and quantity (and delivery date in case of SCML-Standard); failure to meet agreed-upon obligations results in penalties. The ultimate goal for all agents is to maximize their accumulated profit.

SCML's two tracks offer different levels of complexity regarding the negotiation search space and temporal dependencies:

\begin{itemize}
    \item \textbf{OneShot Track:} Products are perishable, meaning profits or losses on a given day are independent of other days, except in the indirect effect of changing partners' future negotiation behavior. Agents act as either buyers or sellers (never both), and the search space is limited by small price and quantity ranges. The research focus here is on \textit{repeated concurrent negotiation} in many-to-many environments.
    \item \textbf{Standard Track:} This track introduces higher complexity by making products nonperishable (carrying storage costs) and allowing agents to negotiate delivery dates. The production graph is deeper, requiring agents to manage both buying and selling simultaneously. This shifts the challenge toward designing strategies for \textit{dependent sequential sets of concurrent negotiations}.
\end{itemize}

\begin{table}[ht!]
	\centering
	\caption{SCML 2025 Results}
	\label{tbl:scml}
	\setlength{\tabcolsep}{2pt}
	\begin{tabular}{l|clr}
		\toprule
        Track & Rank & Agent & Score \\
        \midrule
		OnShot     &    & CautiousOneShotAgent & 1.0915 \\
		           & 1  & CostAverseAgent                          & 1.0896\\
		           & 1  & Rchan                          & 1.0895\\
		           & 1  & AlmostEqualAgent &                       1.0892\\
				   \midrule
		Standard   & 1  & AS0         & 1.009\\
				   & & PenguinAgent               & 0.992\\
		           & 2 & XenoSotaAgent               & 0.960\\
		           & 3 & Ultra Super Miracle Final  Agent Z        & 0.937\\		           
		           
		% \midrule
		% \multicolumn{6}{l}{TUAT: Tokyo University of Agriculture and Technology}\\
		\bottomrule
	\end{tabular}
\end{table}

The performance of the participants is summarized in Table~\ref{tbl:scml}. Notably, the winners of both tracks prioritized domain-specific heuristics over complex opponent modeling (for the second year in a row). 
Since $2020$, all finalists of SCML could outperform the winner of the previous year. This pattern started to break in $2025$ with the winner of SCML-OneShot 2024 (CautiousOneshotAgent) remaining the top performing agent in $2025$ and the winner of SCML-Standard 2024 (PenguinAgent) only outperformed by the winner of SCML-Standard 2025 (AS0). 

Three agents tied for the winner of SCML-Oneshot: 1) CostAverseAgent (developed by Yuzuru Kitamura) which focuses on minimizing the risk of financial loss by prioritizing cost-based decision-making over aggressive profit-seeking. 2) Rchan (developed by Shota Takayama) which is built around a sophisticated dynamic aspiration model and market estimation.  3) AlmostEqualAgent (developed by Kaito Miwa) which utilizes a strategy centered on proportional equity and collaborative surplus sharing to reach agreements quickly.

The winner of SCML-Standard (AS0 developed by Atsunaga et al.) focuses on a risk-balanced production-first strategy that adapts to the complexities of a full supply chain simulation. It employs a conservative negotiation strategy that sets price floors using expected manufacturing costs and storage fees, ensuring that every contract contributes to a positive net profit.

\section{Conclusion}

The 16th International Automated Negotiating Agents Competition (ANAC 2025) has once again demonstrated the evolving complexity of autonomous negotiation in multi-agent systems. By shifting the focus toward sequential multi-deal negotiations in  ANL and deepened supply chain dependencies in the SCML, the competition continues to push the boundaries of how agents handle high-dimensional outcome spaces and temporal uncertainty.

Our analysis of the 2025 results yields several key insights:
\begin{itemize}
\item \textbf{Computational Scalability:} In ANL, the transition to multi-deal scenarios highlighted a critical trade-off between computational overhead and strategic foresight. Success was largely defined by an agent's ability to manage "memory explosion" through tree-search pruning, sampling, or reinforcement learning (e.g., the SAC Agent).
\item \textbf{Heuristics vs. Complexity:} In SCML, the continued success of domain-specific heuristics and risk-averse strategies (e.g., AS0 and CostAverseAgent) suggests that in highly volatile, concurrent markets, robustness and cost-containment often outperform complex opponent modeling.
\item \textbf{Human-Agent Interaction:} The 2025 pilot competition at IJCAI provided a rare benchmark for how autonomous agents perform against human negotiators in survival and trade scenarios, highlighting the need for more "explainable" and human-centric negotiation strategies.
\end{itemize}

Looking forward, the competition aims to further bridge the gap between theoretical frameworks and industrial application. Future iterations of ANAC will likely explore the integration of Large Language Models (LLMs) for more nuanced communication in the Human-Agent Negotiation League planned for 2026. 
ANL will explore the design of a negotiation agent for bilateral negotiation that tries to mislead its opponent. The agent is rewarded for the agreement made in the negotiation, as well as for how well it is able to deceive its opponent. This negotiation-in-the-wild challenge will explore the role of deception and its counter-measures in automated negotiation.

As usual with ANAC since 2010, all agent code is open-sourced and is accessible from the submission website at \url{https://anac.cs.brown.edu}.

As agent-based negotiation moves closer to real-world applications, the lessons learned from ANAC 2025 provide a vital roadmap for developing resilient, efficient, and collaborative autonomous systems.

\appendix

\section*{Ethical Statement}

The Human-Agent Negotiation pilot design received the approval of Ozyegin University, Turkey. Personal information of participants were not kept beyond the announcement of winners. 

\section*{Acknowledgments}

This publication is part of the Vidi project COMBINE (VI.Vidi.203.044) (partly) financed by the Dutch Research Council (NWO). ANAC 2024 was sponsored by AI Journal, NWO, NEC-AIST AI Cooperative Research Laboratory, and Springer AI.

\bibliographystyle{named}
\bibliography{ijcai26}

\end{document}